\documentclass[letterpaper,preprintnumbers,prd,twocolumn,nofootinbib,nobibnotes,showpacs]{revtex4}
\usepackage{amsfonts}
\usepackage{mathrsfs}
\usepackage{epsfig}
\usepackage{graphicx}%
\usepackage{dcolumn}
\usepackage{amsmath}

\makeatletter
\def\btt#1{\texttt{\@backslashchar#1}}%
\DeclareRobustCommand\bblash{\btt{\@backslashchar}}%
\makeatother
\begin{document}

\title{Scalar Field, Four Dimensional Spacetime Volume and the Holographic Dark Energy}
\author{Changjun Gao}
\affiliation{$^{1}$The National Astronomical Observatories,
Chinese Academy of Sciences, Beijing 100012}
\affiliation{{$^{2}$Kavli Institute for Theoretical Physics China,
CAS, Beijing 100190, China }}

\date{\today}

\begin{abstract}
We explore the cosmic evolution of a scalar field which is
identified with the four dimensional spacetime volume. Given a
specific form for the Lagrangian of the scalar field, a new
holographic dark energy model is present. The energy density of
dark energy is reversely proportional to the square of the radius
of the cosmic null hypersurface which is present as a new infrared
cutoff for the Universe.  We find this holographic dark energy
belongs to the phantom dark energy for some appropriate parameters
in order to interpret the current acceleration of the Universe.
\end{abstract}


\maketitle

\section{Introduction}
Scalar fields are of great importance in both physics and
cosmology. In physics, scalar fields are present in
Jordan-Brans-Dicke theory as Jordan-Brans-Dicke scalar
\cite{brans:1961}; in Kaluza-Klein compactification theory as the
radion \cite{csaki:2000}, in the Standard Model of particle
physics as the Higgs boson \cite{higgs:1964}, in the low-energy
limit of the superstring theory as the dilaton \cite{gibbons:1985}
or tachyon \cite{sen:2002} and so on. In cosmology, scalar fields
are investigated as the inflaton \cite{guth:1981} to drive the
inflation of the early Universe and currently as the quintessence
\cite{{ratra:1988},{caldwell:1998},zlatev:1999} or phantom
\cite{caldwell:1999} to drive the acceleration the Universe.

When the canonical scalar field $\phi$ is minimally coupled to the
gravitation, it has the Lagrangian density as follows
 \begin{eqnarray}
\mathscr{L}=\sqrt{-g}\left[\frac{1}{2}\nabla_{\mu}\phi\nabla^{\mu}\phi+V\left(\phi\right)\right]\;,
\end{eqnarray}
where $V\left(\phi\right)$ is the scalar potential. The scalar
field may originate from the extra space dimensions, for example,
 in Kaluza-Klein compactification theory as the radion \cite{csaki:2000}, in the low-energy
limit of the superstring theory as the dilaton \cite{gibbons:1985}
 or tachyon \cite{sen:2002}.

 Now we want to know whether we can construct the scalar field
 from the four dimensional spacetime geometry. In General Relativity,
the dynamical variable is the metric tensor $g_{\mu\nu}$ from
which we can construct the Ricci scalar ${R}$, the four
dimensional volume,
\begin{eqnarray}
\texttt{V}=\int d^4 x\sqrt{-g}\;,
\end{eqnarray}
and various scalar quantities (for example, $R^2$,
$R_{\mu\nu}R^{\mu\nu}$,
$R_{\mu\nu\delta\lambda}R^{\mu\nu\delta\lambda}$ and so on).

However, among the vast scalars, it is uniquely the four
dimensional volume that has no derivatives with respect to the
spacetime coordinates. So in order to obtain an equation of motion
up to the second order of derivatives, if and only if we assume
the scalar field $\phi$ is the function of four dimensional
spacetime volume
\begin{eqnarray}
\phi=\phi\left(\texttt{V}\right)\;.
\end{eqnarray}

For simplicity, we identify $\phi$ with the four dimensional
volume,
\begin{eqnarray}
\phi\equiv\texttt{V}\;.
\end{eqnarray}
We define the kinetic term of the scalar field as $U$:
\begin{eqnarray}
\label{eq:U} U\equiv
-\frac{1}{2}\nabla_{\mu}\phi\nabla^{\mu}\phi\;.
\end{eqnarray}
In the following, we shall explore the Lagrangian density as
follows
\begin{eqnarray}
\label{eq:LL} \mathscr{L}=K\left(U\right)\sqrt{-g}\;,
\end{eqnarray}
with $K$ an arbitrary function of $U$. It is similar to the pure
K-essence theory \cite{muk:2000}.

Now the total action in the presence of other matter sources is
given by
\begin{eqnarray}
S=\int d^4
x\sqrt{-g}\left[\frac{R}{16\pi}+K\left(U\right)\right]+S_{m}\;.
\end{eqnarray}
In the next, we will investigate the cosmic evolution of this
scalar field in the presence of other matter sources which include
matter (baryon matter and dark matter) and radiation. We find the
scalar field behaves as a holographic dark energy
\cite{Li:2004,holo:2009,cai:2007,wei:2008,gao:2009,rongjia:2011,fischler:1998}
for the specific form of $K(U)\propto \sqrt[3]{1/U}$.

The paper is organized as follows. In section II, we shall
calculate the four dimensional spacetime volume of the
Friedmann-Robertson-Walker (FRW) Universe. To this end, we rewrite
the FRW metric from the homogenous and isotropic coordinate system
to the Schwarzschild coordinate system. This is motivated by
Faraoni's recent paper \cite{faraoni:2011} where the dynamics of
particle, event, and apparent horizons of FRW Universe is studied.
In section III, we construct the Lagrangian $K(U)$ for the
holographic dark energy. In section IV, we investigate the cosmic
evolution of the scalar field and find it is a phantom dark energy
for some appropriate parameters in order to interpret the current
acceleration of the Universe. Section V gives the conclusion and
discussion.
\section{four dimensional volume}
In this section, let's calculate the four dimensional spacetime
volume of our Universe. We mainly follow Faraoni's recent paper
\cite{faraoni:2011}. Consider spatially flat
Friedmann-Robertson-Walker (FRW) Universe which has the metric
\begin{eqnarray}
 ds^2=-dt^2+a^2\left(dr^2+r^2d\Omega^2\right)\;,
\end{eqnarray}
where $a(t)$ is the scale factor. The coordinate system $(t, r,
\theta, \varphi)$ is named after \emph{homogenous and isotropic
coordinate system}. In order to calculate the four dimensional
volume, we had better rewrite the metric in the
\emph{Schwarzschild coordinate system}. Therefore, we introduce
the physical space variable $\bar{r}$ by

\begin{eqnarray}
\bar{r}\equiv a\left(t\right)r\;.
\end{eqnarray}
Then the metric becomes
\begin{eqnarray}
\label{metric}
ds^2=-\left(1-H^2\bar{r}^2\right)dt^2-2H\bar{r}dtd\bar{r}+d\bar{r}^2+\bar{r}^2d\Omega^2\;,
\end{eqnarray}
where
\begin{eqnarray}
H\equiv \frac{1}{a}\frac{da}{dt}\;,
\end{eqnarray}
is the Hubble parameter.
 The coordinate system of
$(t, \bar{r}, \theta, \varphi)$ is not orthogonal. We can
eliminate the coefficient of $dtd\bar{r}$ by introducing a new
time coordinate, $T$. The form of Eq.~(\ref{metric}) suggests we
set
\begin{equation}
\label{FF}
dT=\frac{1}{F\left(t,\bar{r}\right)}\left[dt+\frac{\bar{r}H}{1-H^2\bar{r}^2}d\bar{r}\right],
\end{equation}
where $F(t,\bar{r})$ is a perfect differential factor and it
always exists. $F(t,\bar{r})$ solves the equation
\begin{equation}
\label{FFF}
\frac{\partial}{\partial\bar{r}}\left(\frac{1}{F\left(t,\bar{r}\right)}\right)=\frac{\partial}{\partial
t}\left[\frac{\bar{r}H}{F\left(1-H^2\bar{r}^2\right)}\right].
\end{equation}
Then the metric Eq.~(\ref{metric}) can be written as
\begin{equation}
\label{metric:sch}
ds^2=-\left(1-H^2\bar{r}^2\right)F^2dT^2+\frac{1}{1-H^2\bar{r}^2}d\bar{r}^2
+\bar{r}^2d\Omega^2\;,
\end{equation}
where $H$ and $F$ become now the functions of variables $T$ and
$\bar{r}$. This is the FRW metric in the Schwarzschild coordinate
system. For de Sitter Universe, we have $a=e^{Ht}$ with $H$ a
constant. Then Eq.~(\ref{FF}) or Eq.~(\ref{FFF}) tell us we may
put $F=1$ and Eq.~(\ref{metric:sch}) reduces exactly to the
well-known form:
\begin{equation}
\label{metric:sch0}
ds^2=-\left(1-H^2\bar{r}^2\right)dT^2+\frac{1}{1-H^2\bar{r}^2}d\bar{r}^2
+\bar{r}^2d\Omega^2\;.
\end{equation}
For FRW Universe, $H$ and $F$ are the functions of variables $T$
and $\bar{r}$. It is apparent the Universe is isotropic but not
homogeneous in the Schwarzschild coordinate system. From
Eq.~(\ref{metric:sch}) we see there exists a horizon with the
physical radius
\begin{equation}
\bar{r}_H=H^{-1}\;,
\end{equation}
at which $g_{00}=0$ and $g_{11}=\infty$. It is usually called the
Hubble horizon or dynamical apparent horizon \cite{gong:07}. The
Hubble-redshift relation is given by
\begin{equation}
\label{eq:A6} v=H\bar{r}\;,
\end{equation}
where $v$ can be interpreted as the receding velocity of galaxies
or cluster of galaxies. Substituting Eq.~(\ref{eq:A6}) into
Eq.~(\ref{metric:sch}), we obtain
\begin{equation}
\label{metric:sch1}
ds^2=-\left(1-v^2\right)F^2dT^2+\frac{1}{1-v^2}d\bar{r}^2
+\bar{r}^2d\Omega^2\;.
\end{equation}
So the receding velocity $v$ approaches the speed of light on the
Hubble horizon $\bar{r}_H=H^{-1}$. But we find in the next the
Hubble horizon is actually not a null hypersurface because it does
not obey the equation for a null hypersurface.

To show this point, we solve the equation of null hypersurface
\begin{equation}
\label{eq:null00} f\left(x^{\mu}\right)=0 \;.
\end{equation}
Taking into account the spherically symmetric property of the
Universe, the null hypersurface should have the form
\begin{equation}
\label{eq:null0} f\left(T,\bar{r}\right)=0 \;,
\end{equation}
which is determined by the definition
\begin{equation}
\label{eq:null} g^{\mu\nu}\frac{\partial f}{\partial
x^{\mu}}\frac{\partial f}{\partial x^{\nu}}=0 \;.
\end{equation}
Eq.~(\ref{eq:null}) can be rewritten as follows
\begin{equation}
\label{eq:nulle}
-\left(1-H^2\bar{r}^2\right)^{-1}F^{-2}\frac{\partial f}{\partial
T}\frac{\partial f}{\partial
T}+\left(1-H^2\bar{r}^2\right)\frac{\partial f}{\partial
\bar{r}}\frac{\partial f}{\partial\bar{r}}=0 \;.
\end{equation}
From Eq.~(\ref{eq:null0}) we obtain
\begin{equation}
\label{eq:nn} \bar{r}_N=\bar{r}_N\left(T\right)\;,
\end{equation}
where $\bar{r}_N$ represents the physical radius of the null
hypersurface. Substituting Eq.~(\ref{eq:nn}) into
Eq.~(\ref{eq:nulle}), we obtain the equation for the null
hypersurface
\begin{equation}
\frac{\partial\bar{r}_N}{\partial
T}=F\left(1-H^2\bar{r}_N^2\right)\;.
\end{equation}
Turn to the cosmic time $t$, we arrive at (from Eq.~(10))
\begin{equation}
\frac{\partial\bar{r}_N}{\partial t}=1-H\bar{r}_N\;.
\end{equation}
Taking account of $H$ as the function of cosmic time $t$, we
recognize that above equation describes the evolution of physical
radius $\bar{r}_N$ of the null hypersurface with cosmic time $t$.
We note that this null hypersurface is different from the particle
horizon \cite{part}
\begin{eqnarray}
r_P&=&a\left(t\right)\int_0^t \frac{1}{a\left(t\right)}dt\;,
\end{eqnarray}
the event horizon \cite{part}
\begin{eqnarray}
r_E&=&a\left(t\right)\int_{t}^{\infty}
\frac{1}{a\left(t\right)}dt\;,
\end{eqnarray}
and the apparent horizon \cite{gong:07} (for spatially flat
Universe)
\begin{eqnarray}
r_A&=&\frac{1}{H}\;.
\end{eqnarray}
The difference comes from the fact $\bar{r}_N$ is defined in not
the isotropic and homogenous coordinate system, but the
Schwarzschild coordinate system. We focus on the four dimensional
volume within the null hypersurface $\bar{r}_N$:
\begin{eqnarray}
\phi &=&\int d^4 x\sqrt{-g}=\int 4\pi \bar{r}^2 d\bar{r}dt
\nonumber\\&=&\int_{0}^{\bar{r}_N} d\bar{r}\int_{0}^{t}dt\cdot
4\pi \bar{r}^2 =\int_{0}^{t}{\frac{4}{3}\pi\bar{r}_N^3}dt\;.
\end{eqnarray}
In particular, for de Sitter space (Eq.~\ref{metric:sch0}), we
have
\begin{eqnarray}
\bar{r}_N=r_E=r_A=\frac{1}{H}\;,
\end{eqnarray}
and
\begin{eqnarray}
\phi&=&{\frac{4}{3}\pi}H^{-3}t\;.
\end{eqnarray}
\section{Holographic dark energy}
In section II, we derived the four dimensional volume for the
Universe. In this section, we shall derive the Lagrangian density
of holographic dark energy using the four dimensional volume.
According to the holographic dark energy model, the density of
dark energy should inversely proportional to the square of some
horizon, for example, the event horizon \cite{Li:2004}, the
particle horizon \cite{fischler:1998}, the four dimensional Ricci
radius \cite{gao:2009}, the three dimensional Ricci radius
\cite{rongjia:2011} and so on. In the next, lets's investigate the
case for the null hypersurface. Then the density of dark energy is
given by
\begin{eqnarray}
\label{holo} {\bar{\rho}}=\frac{\alpha}{\bar{r}_N^2}\;,
\end{eqnarray}
where $\alpha$ is a positive constant. Since $\bar{r}_N$ is
different from $r_P$, $r_E$ and $r_A$, it turns out to be a new
infrared cutoff for the Universe. It is easy to find the kinetic
energy $U$ from Eq.~(\ref{eq:U})
\begin{eqnarray}
\label{NNN} U\equiv
-\frac{1}{2}\nabla_{\mu}\phi\nabla^{\mu}\phi=\frac{8}{9}\pi^2\bar{r}_N^6\;.
\end{eqnarray}
The pressure derived from the Lagrangian Eq.~(\ref{eq:LL}) is
given by
\begin{eqnarray}
\bar{p} &=&\frac{1}{8\pi}\cdot\frac{1}{6a^2}\cdot\frac{\partial
\mathscr{L}}{\partial a}
\nonumber\\&=&\frac{1}{8\pi}\cdot\frac{1}{6a^2}\cdot\frac{\partial
{\left(K a^3\right)}}{\partial a} \;.
\end{eqnarray}
 On the other hand, the energy-momentum conservation equation,
\begin{eqnarray}
\bar{p} &=&-\bar{\rho}-\frac{1}{3}\frac{\partial
\bar{\rho}}{\partial \ln a} \;,
\end{eqnarray}
can be rewritten as
\begin{eqnarray}
-\bar{\rho}-\frac{1}{3}\frac{\partial \bar{\rho}}{\partial \ln a}
=\frac{1}{8\pi}\cdot\frac{1}{6a^2}\cdot\frac{\partial {\left(K
a^3\right)}}{\partial a} \;.
\end{eqnarray}
Substituting Eq.~(\ref{holo}) into above equation, we obtain
\begin{eqnarray}
K\left(U\right)=\sqrt[3]{\frac{\beta}{U}}\;,
\end{eqnarray}
with
\begin{eqnarray}
\beta\equiv-\frac{32^3}{9}\alpha^3\pi^5\;.
\end{eqnarray}
So the Lagrangian density for the holographic dark energy is
\begin{eqnarray}
\label{eq:LLL}
\mathscr{L}=\sqrt[3]{\frac{\beta}{U}}\cdot\sqrt{-g}\;.
\end{eqnarray}
The reason for $\sqrt[3]{{1}/{U}}$ could also be understood from
the dimensional analysis. Eq.~(2) and Eq.~(3) tell us the
dimension of $\phi$ is $l^{4}$. So the dimension of $U$
(Eq.~(\ref{NNN})) is $l^6$. We conclude the dimension of
$\sqrt[3]{{1}/{U}}$ is $l^{-2}$ which is the same as Ricci scalar.
$\beta$ is present as a dimensionless constant.
\section{cosmic evolution}

In this section, we investigate the cosmic evolution of this dark
energy proposal in detail. We model all other matter sources
present in the Universe as the perfect fluids. These matter
sources can be baryon matter, dark matter and relativistic matter
and so on. We assume there is no interaction between the scalar
field and other matter fields. Then the Friedmann equation is
given by
\begin{eqnarray}
\label{eq:IR}
3H^2&=&8\pi\left(\rho+\frac{\alpha}{\bar{r}_N^2}\right)\;.
\end{eqnarray}
$\bar{r}_N$ is determined by
\begin{equation}
\label{eq:main} H\frac{\partial\bar{r}_N}{\partial \ln
a}=1-H\bar{r}_N\;.
\end{equation}

For the present-day Universe, we have
\begin{eqnarray}
\label{eq:pfe} 3H_0^2=8\pi\rho_0\;,
\end{eqnarray}
where $H_0$ and $\rho_0$ are the present-day Hubble parameter and
the present-day total energy density. Divided Eq.~(\ref{eq:IR}) by
Eq.~(\ref{eq:pfe}) and put
\begin{eqnarray}
&& h=\frac{H}{H_0}\;,\ \ \
\Omega_{m0}=\frac{\rho_{m0}}{\rho_0}\;,\ \ \
\Omega_{r0}=\frac{\rho_{r0}}{\rho_0}\;,\nonumber\\&&
\eta=\frac{8\pi}{3}\alpha\;,\ \ \ u=H_0\bar{r}_N\;,
\end{eqnarray}
where $\Omega_{m0}$ and $\Omega_{r0}$ are the relative density of
the dark matter and the radiation, respectively. The main
equations are reduced to
\begin{eqnarray}
\label{eq:h}
{h}^2&=&\frac{\Omega_{m0}}{a^{3}}+\frac{\Omega_{r0}}{a^{4}}+\frac{\eta}{u^{2}}\;,\nonumber\\
h\frac{\partial u}{\partial\ln a}&=&1-hu\;.
\end{eqnarray}
Let
\begin{eqnarray}
x\equiv \ln a\;.
\end{eqnarray}
We obtain
\begin{eqnarray}
\label{eq:h}
{h}^2&=&{\Omega_{m0}}{e^{-3x}}+{\Omega_{r0}}{e^{-4x}}+\frac{\eta}{u^{2}}\;,\nonumber\\
h\frac{\partial u}{\partial x}&=&1-hu\;.
\end{eqnarray}

It is hard to solve these differential equations analytically.
Let's set up an autonomous system to study the evolution of the
Universe. Following Ref.~\cite{cope:1999}, we introduce the
following dimensionless quantities
\begin{eqnarray}
X\equiv\sqrt{\Omega_{m0}}\cdot\frac{e^{-3x/2}}{h}\;,\ \ \
Y\equiv\sqrt{\Omega_{r0}}\cdot\frac{e^{-2x}}{h}\;.
\end{eqnarray}
Then $X^2$ and $Y^2$ represent the density parameters for the
matter and radiation, respectively. The main equations can be
written in the following autonomous form
\begin{eqnarray}
\label{eq:auto}
\frac{dX}{dx}&=&-\frac{3}{2}X-\frac{1}{2}X\left[-3X^2-4Y^2-2\left(1-X^2-Y^2\right)\right.\nonumber\\&&\left.
\cdot\left(\frac{\sqrt{1-X^2-Y^2}}{\sqrt{\eta}}-1\right)\right]\;,\\
\frac{dY}{dx}&=&-{2}Y-\frac{1}{2}Y\left[-3X^2-4Y^2-2\left(1-X^2-Y^2\right)\right.\nonumber\\&&\left.
\cdot\left(\frac{\sqrt{1-X^2-Y^2}}{\sqrt{\eta}}-1\right)\right]\;,
\end{eqnarray}
together with a constraint equation
\begin{eqnarray}
X^2+Y^2+\Omega_{\phi}=1\;.
\end{eqnarray}
Here $\Omega_{\phi}$ stands for the density parameter of
holographic dark energy. The equation of state $w$ of dark energy
is found to be
\begin{eqnarray}
w=-1+\frac{2}{3}\cdot\left(\frac{\sqrt{1-X^2-Y^2}}{\sqrt{\eta}}-1\right)\;.
\end{eqnarray}
The deceleration parameter $q$ of the Universe is given by
\begin{eqnarray}
q&=&-1-\frac{1}{2}\left[-3X^2-4Y^2-2\left(1-X^2-Y^2\right)\right.\nonumber\\&&\left.
\cdot\left(\frac{\sqrt{1-X^2-Y^2}}{\sqrt{\eta}}-1\right)\right]\;.
\end{eqnarray}

\begin{table*}[t]
\begin{center}
\begin{tabular}{|c|c|c|c|c|c|c|c|c|}
Name &  $X$ & $Y$ & Existence & Stability & $\Omega_\phi$
 & $w$ \\
\hline \hline (a1) & 0 & 0 & $0<\eta<\frac{1}{9}$& Unstable node
&   1 & $-\frac{5}{3}+\frac{2}{3\sqrt{\eta}}$ \\
\hline (a2) & 0 & 0 &
$\frac{1}{9}<\eta<\frac{4}{25}$ & Saddle point & $1$ &$-\frac{5}{3}+\frac{2}{3\sqrt{\eta}}$\\
\hline (a3) & 0&0 &
$\frac{4}{25}<\eta$ & Stable node& $1$ &$-\frac{5}{3}+\frac{2}{3\sqrt{\eta}}$\\
\hline (b) & $1$  &0& $\eta>0$ & Saddle point & $0$ & $-\frac{5}{3}$  \\
\hline (c) & $0$  &1& $\eta>0$ & Unstable node & $0$ & $-\frac{5}{3}$  \\
\hline (d) & $0$  &$\sqrt{1-9\eta}$ & $0<\eta<\frac{1}{9}$ & Saddle point & $9\eta$ & $\frac{1}{3}$  \\
\hline (e) & $\frac{\sqrt{4-25\eta}}{2}$&0 & $0<\eta<\frac{4}{25}$ & Stable node & $\frac{25}{4}\eta$ &$0$\\
\hline
\end{tabular}
\end{center}
\caption[crit]{The properties of the critical points for the
autonomous system.} \label{crit0}
\end{table*}

In TABLE I, we present the properties of the critical points for
different values of $\eta$. The points (a1,\ a2,\ a3) correspond
to the dark energy dominated epoch and the point (a3) are stable.
In this epoch, the equation of state for dark energy is
$-\frac{5}{3}+\frac{2}{3\sqrt{\eta}}$. The point (b) corresponds
to the matter dominated epoch and it is a saddle point. The point
(c) corresponds to the radiation dominated epoch and it is an
unstable node. In both epoches, the equation of state for dark
energy is $-\frac{5}{3}$. Point (d) and point (e) correspond to
the track solution that the dark energy tracks the background
energy sources (matter and radiation).

The present-day matter density parameter $\Omega_{m0}$ and
radiation density parameter $\Omega_{r0}$ have been obtained by
Komatsu et al. \cite{komatsu:2009} from a combination of baryon
acoustic oscillation, type Ia supernovae and WMAP5 data at a
$95\%$ confidence limit, $\Omega_{m0}=0.25$ and
$\Omega_{r0}=8.1\cdot10^{-5}$. So in the following discussions, we
will adopt these values.
\begin{figure}
\includegraphics[width=7.8cm]{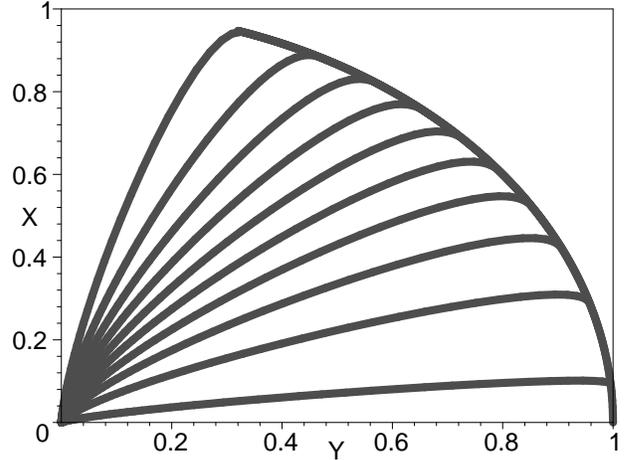}
\ \ \ \ \ \ \ \ \ \caption{The phase plane for $\eta=0.9$. The
point ($X=0,\ Y=0$) corresponds to the dark energy dominated
epoch. It is stable and thus an attractor. The point ($X=0,\ Y=1$)
corresponds to the radiation dominated epoch at which the dark
energy has the equation of state $-5/3$. The point ($X=1,\ Y=0$)
corresponds to the matter dominated epoch at which the dark energy
has also the equation of state $-5/3$. These trajectories show
that the Universe always evolves from the radiation dominated
epoch to the dark energy dominated epoch.}\label{fig:expon}
\end{figure}

In Fig.~1, we plot the phase portraits for $\eta=0.9$ with vast
initial conditions. For $\eta=0.9$, we have three critical points,
namely, point (0, 0), (0, 1) and (1, 0). The point (0, 0)
corresponds to the dark energy dominated epoch and the circled arc
($X^2+Y^2=1$) corresponds to the matter and radiation co-dominated
epoch. The point (0,\ 0) is stable and thus an attractor. The
point (1, 0) is the matter dominated epoch and it is a saddle
point. The point (0, 1) is the radiation dominated epoch and it is
unstable. Since this point is unstable, we conclude that the
Universe always evolves from the radiation dominated epoch to the
dark energy dominated epoch shown by the portrait.

\begin{figure}
\includegraphics[width=8.4cm]{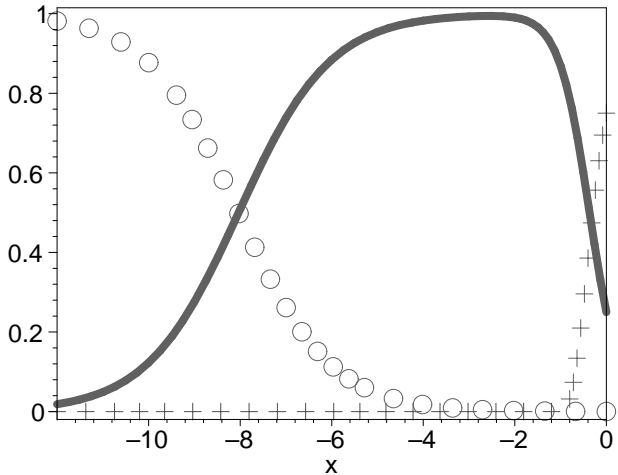}
\ \ \caption{The evolution of density fractions for radiation
($Y^2$)
  (circled line), matter ($X^2$) (solid line) and dark energy ($\Omega_{\phi}$) (crossed line), for the
 value $\eta=0.9$.}
\label{fig:dens}
\end{figure}

In Fig.~\ref{fig:dens}, we plot the evolution of density fractions
for radiation, matter and holographic dark energy, for the
parameter $\eta=0.9$. The solid line represents the density
fraction of the matter. The circled line and crossed line
represent the density fraction for the radiation and dark energy,
respectively.

\begin{figure}
\includegraphics[width=8.9cm]{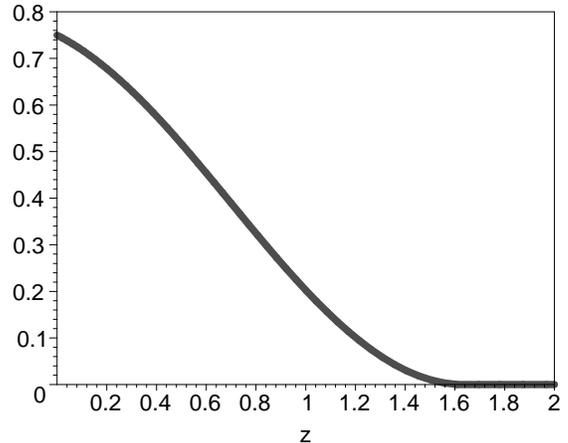}
\\
\caption{The evolution of the dimensionless energy density
$\eta/u^2$ of holographic dark energy with redshift. The energy
density of dark energy approaches zero at redshifts greater than
$1.6$.} \label{fig:de}
\end{figure}

In Fig.~\ref{fig:de}, we plot the evolution of the dimensionless
energy density $\eta/u^2$ of holographic dark energy with
redshift. Since the density of dark energy approaching zero at
redshifts greater than $1.6$, this shows that dark energy should
not play a key role in the history of structure formation. Also,
the coincidence problem \cite{coin:2000} is greatly relaxed
because the scalar field emerges very recently.

\begin{figure}
\includegraphics[width=8.5cm]{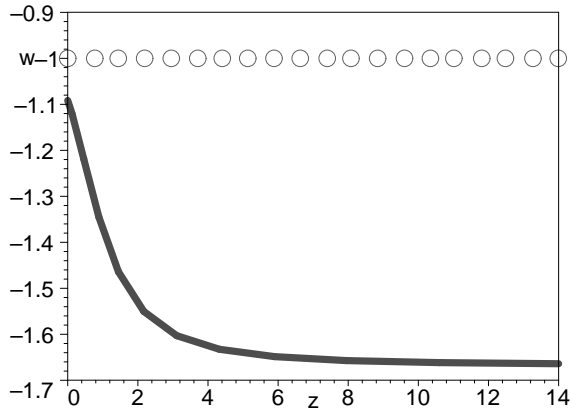}
\\
\caption{The equation of the state of the dark energy. Since the
equation of state is always smaller than $-1$, it belongs to the
phantom dark energy models. The present-day equation of state is
$-1.1$ which is consistent with observations.} \label{fig:wz}
\end{figure}

In Fig.~\ref{fig:wz}, we plot the equation of state $w$ for dark
energy. Since the equation of state is always smaller than $-1$,
the dark energy belongs to the phantom dark energy models. For the
present universe, the equation of state is $-1.1$. This is
consistent with observations \cite{komatsu:2009}.

\begin{figure}
\includegraphics[width=8.4cm]{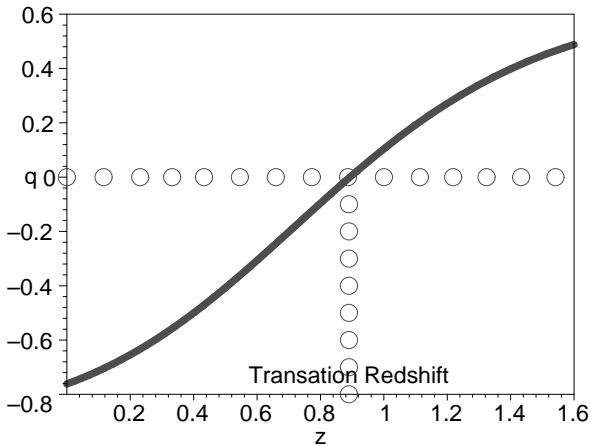}
\\
\caption{The deceleration parameter $q$ with the redshift. The
Universe speeds up around redshift $z_{T}=0.86$ which is not
inconsistent with astronomical observations.} \label{fig:qz}
\end{figure}

In Fig.~\ref{fig:qz}, we plot the deceleration parameter $q$ for
the Universe. Fig.~\ref{fig:qz} tells us the Universe speeds up
around redshift $z_{T}=0.86$ which is not inconsistent with
astronomical observations.

\section{conclusion and discussion}

Scalar fields may originate from the extra space dimensions, for
example, in Kaluza-Klein compactification theory as the radion
\cite{csaki:2000}, in the low-energy limit of the superstring
theory as the dilaton \cite{gibbons:1985} or tachyon
\cite{sen:2002}. In this paper, we propose the scalar field
originates from the four dimensional spacetime geometry. In order
that the theory leads to the second order differential equations,
one should let $\phi$ be the function of the four dimensional
volume. For simplicity, we identify $\phi$ with the four
dimensional volume. Motivated by Faraoni's work
\cite{faraoni:2011}, we rewrite the FRW metric in the
Schwarzschild coordinate system and then calculate the four
dimensional volume.

For the specific form of Lagrangian $\propto 1/\sqrt[3]{U}$, the
model gives the holographic dark energy where the density of dark
energy is inversely proportional to the square of the radius of
null hypersurface. We note that this null hypersurface is
different from the Hubble horizon, the event horizon, the particle
horizon and the apparent horizon. In order to interpret the
current acceleration of the Universe, this holographic dark energy
belongs to the phantom dark energy.

Is this holographic dark energy consistent with the solar system
tests on General Relativity? The answer is \emph{yes}. The reason
could be understood as follows. The gravitational field in the
solar system can be very well described by the Schwarzschild
solution
\begin{eqnarray}
ds^2=-\left(1-\frac{2M}{r}\right)dt^2+\left(1-\frac{2M}{r}\right)^{-1}dr^2+r^2d\Omega^2\;,
\end{eqnarray}
where $M$ is the mass of the Sun. The four dimensional volume of
the spacetime is
\begin{eqnarray}
\texttt{V}=\int d^4 x\sqrt{-g}=\int_{0}^{L_{\infty}}{4}\pi
r^2dr\int_0^{t}dt=\frac{32}{3}\pi L_{\infty}^3 t\;.
\end{eqnarray}
Here $L_{\infty}$ stands for the maximum scale in the Universe
which can be taken as the Universe scale $L_{\infty}=\bar{r}_N$.
Then the Lagrangian (Eq.~(\ref{eq:LLL})) becomes
\begin{eqnarray}
\label{AAA}
\mathscr{L}\propto\sqrt{-g}\cdot\frac{1}{\bar{r}_{N}^2}\;.
\end{eqnarray}
We know from the dimensional analysis that the Ricci scalar
$R\simeq{L^{-2}}$ with $L$ some scale within the solar system. It
is apparent that
\begin{eqnarray}
 \frac{1}{\bar{r}_{N}^2}\ll \frac{1}{L^2}\;,
\end{eqnarray}
in the solar system. So the Lagrangian (Eq.~(\ref{AAA})) can be
safely neglected compared to the Einstein-Hilbert action. In other
words, the gravitational field in the solar system can be very
well described by the Schwarzschild solution.

\acknowledgments

This work is supported by the National Science Foundation of China
under the Key Project Grant 10533010, Grant 10575004, Grant
10973014, and the 973 Project (No. 2010CB833004).

\newcommand\ARNPS[3]{~Ann. Rev. Nucl. Part. Sci.{\bf ~#1}, #2~ (#3)}
\newcommand\AL[3]{~Astron. Lett.{\bf ~#1}, #2~ (#3)}
\newcommand\AP[3]{~Astropart. Phys.{\bf ~#1}, #2~ (#3)}
\newcommand\AJ[3]{~Astron. J.{\bf ~#1}, #2~(#3)}
\newcommand\APJ[3]{~Astrophys. J.{\bf ~#1}, #2~ (#3)}
\newcommand\APJL[3]{~Astrophys. J. Lett. {\bf ~#1}, L#2~(#3)}
\newcommand\APJS[3]{~Astrophys. J. Suppl. Ser.{\bf ~#1}, #2~(#3)}
\newcommand\JHEP[3]{~JHEP.{\bf ~#1}, #2~(#3)}
\newcommand\JCAP[3]{~JCAP. {\bf ~#1}, #2~ (#3)}
\newcommand\LRR[3]{~Living Rev. Relativity. {\bf ~#1}, #2~ (#3)}
\newcommand\MNRAS[3]{~Mon. Not. R. Astron. Soc.{\bf ~#1}, #2~(#3)}
\newcommand\MNRASL[3]{~Mon. Not. R. Astron. Soc.{\bf ~#1}, L#2~(#3)}
\newcommand\NPB[3]{~Nucl. Phys. B{\bf ~#1}, #2~(#3)}
\newcommand\PLB[3]{~Phys. Lett. B{\bf ~#1}, #2~(#3)}
\newcommand\PRL[3]{~Phys. Rev. Lett.{\bf ~#1}, #2~(#3)}
\newcommand\PR[3]{~Phys. Rep.{\bf ~#1}, #2~(#3)}
\newcommand\PRD[3]{~Phys. Rev. D{\bf ~#1}, #2~(#3)}
\newcommand\RMP[3]{~Rev. Mod. Phys.{\bf ~#1}, #2~(#3)}
\newcommand\SJNP[3]{~Sov. J. Nucl. Phys.{\bf ~#1}, #2~(#3)}
\newcommand\ZPC[3]{~Z. Phys. C{\bf ~#1}, #2~(#3)}
 \newcommand\IJGMP[3]{~Int. J. Geom. Meth. Mod. Phys.{\bf ~#1}, #2~(#3)}
  \newcommand\GRG[3]{~Gen. Rel. Grav.{\bf ~#1}, #2~(#3)}
\newcommand\IJMPA[3]{~Int. J. Mod. Phy. A{\bf ~#1}, #2~(#3)}
\newcommand\CTP[3]{~Commun. Theor. Phys.{\bf ~#1}, #2~(#3)}
\newcommand\MPLA[3]{~Mod. Phys. Lett. A.{\bf ~#1}, #2~(#3)}

\end{document}